\documentclass[12pt]{iopart}
\usepackage{iopams}
\usepackage{graphics}
\newcommand{\dt}{\frac{d}{d t}}
\newcommand{\da}{\rho_A}
\newcommand{\df}{\rho_F}
\newcommand{\lan}{\langle}
\newcommand{\ran}{\rangle}
\newcommand{\bd}{{\bf d}}
\newcommand{\pl}{\hat{\bf d}}
\newcommand{\E}{\hat{\bf E}}
\newcommand{\EE}{{\bf E}}
\newcommand{\G}{\textbf{\textit{G}}}

\newcommand{\bra}{{\bf r}}
\newcommand{\bk}{{\bf k}}
\newcommand{\bks}{{{\bf k},s}}
\newcommand{\crn}{\hat{a}_{\bks}^{\dagger}}
\newcommand{\ann}{\hat{a}_{\bks}}

\newcommand{\ar}{\hat{\sigma}^+}
\newcommand{\al}{\hat{\sigma}^-}
\newcommand{\ee}{\hat{\epsilon}_\bks}

\newcommand{\rf}[1]{(\ref{#1})}
\begin{document}

\title[Emission spectra and intrinsic bistability]{Emission spectra and intrinsic optical bistability in a two-level medium}

\author{M. G. Gladush$^{1,3}$, D. V. Kuznetsov$^2$, A. A. Panteleev$^{2,3}$,\\
Vl. K. Roerich$^{2,3}$}

\address{$^1$ Institute for Spectroscopy of the Russian Academy of Sciences, Fizicheskaya Str. 5, Troitsk, Moscow region, 142190 Russia}
\address{$^2$ State research center of Russian Federation Troitsk Institute for Innovation and Fusion Research, Pushkovykh Str., vlad. 12, Troitsk, Moscow region, 142190 Russia}
\address{$^3$ Vladimir State University, Gorkogo Str. 87, Vladimir, 600000 Russia}
\ead{mglad@isan.troitsk.ru}
\begin{abstract}
Scattering of resonant radiation in a dense two-level medium is studied theoretically with account for local field effects and renormalization of the resonance frequency. Intrinsic optical bistability is viewed as switching between different spectral patterns of fluorescent light controlled by the incident field strength. Response spectra are calculated analytically for the entire hysteresis loop of atomic excitation. The equations to describe the non-linear interaction of an atomic ensemble with light are derived from the Bogolubov-Born-Green-Kirkwood-Yvon hierarchy for reduced single particle density matrices of atoms and quantized field modes and their correlation operators. The spectral power of scattered light with separated coherent and incoherent constituents is obtained straightforwardly within the hierarchy. The formula obtained for emission spectra can be used to distinguish between possible mechanisms suggested to produce intrinsic bistability in experiments.
\end{abstract}

\submitto{\jpb}
\maketitle

\section{Introduction}
\label{intro}
It has long been known that optical properties of dense atomic ensembles or complex materials may be very different from those exhibited by independent atoms. Some of these systems can produce two different output signals responding to the same input intensity from a driving laser. For the entire range of applied laser intensities the output function would form a hysteresis loop. When no external feedback is required for this phenomenon to occur it is recognized as intrinsic optical bistability (IOB). Experimentally IOB has been observed in the optical response from rare earth ions and ion pairs in glasses and crystallines \cite{Hehlen_94,Hehlen_96,Hehlen_98,Hehlen_99,Hehlen_2000,Gudel_2000,Redmond_2003,Ward_2007}.
However, its physical nature is still subject to debate. Most theoretical models suggest IOB originates from the fact that the transition frequency of a light emitter $\omega_A$ in a driven system is somehow renormalized and takes the effective value $\bar{\omega}_A=\omega_A-\zeta W$. The latter is a linear function of inversion $W$ where the renormalization constant $\zeta$ follows from the coupling mechanism between the light emitter and its surrounding. Since $W$ is $\bar{\omega}_A$ dependent, it provides the system with an intrinsic feedback and nonlinearity and, therefore, produces bistability in the steady state solution. Most commonly such renormalization is introduced through near dipole-dipole interactions (NDD) between emitters \cite{Bowden_2000_OC,Crenshaw_08,Novitsky_10}. Alternatively, it can be achieved by considering model interactions for emitter pairs \cite{Guillot_2001,Guillot_2002PRB,Guillot_2002PRA}, coupling between a single emitter and the local vibration mode in its host medium \cite{Messina_2003,Messina_2004} or in an interacting charge-transfer exciton system \cite{ZhuLi_2000}. But, certainly, the IOB community has always been open to other ideas.

Originally IOB was predicted by Bowden with co--workers for a dense ensemble of two-level atoms \cite{Bowden_79,Bowden_84,Bowden_86_PRA,Bowden_86_Opt}. Currently, in a simple semiclassical view to the problem, $\bar{\omega}_A$ is a result of the famous Lorentz local-field correction (LFC). It states that in a dense homogeneous ensemble exposed to a laser light each atom is actually driven by an effective or local field strength
\begin{equation}
  \EE_{L}=\EE_M+\frac{4\pi}{3}\mathbf{P},
  \label{LFC}
\end{equation}
where $\EE_M$ is the macroscopic field in the medium commonly approximated as the laser field and $\mathbf{P}$ is the macroscopic polarization. If one substitutes this condition to the optical Bloch equations this corrected field can be manipulated to give the effective $\bar{\omega}_A=\omega_A-\zeta_{L} W$ for which $\zeta_{L}=(4\pi/3\hbar)N|\mathbf{d}|^2$ is the Lorentz frequency or NDD interaction parameter. It depends on the number density of atoms $N$ and the matrix element of the atomic transition dipole moment $\mathbf{d}$. In the literature $\zeta_{L}W$ is often referred to as Bowden-Lorentz redshift \cite{Crenshaw_08}. Thus it must be widely understood that the Bowden type excitation-dependent renormalization of the resonance frequency follows straightforwardly from the locally effective Rabi frequency $\bar{\Omega}=\mathbf{d}\cdot\EE_{L}/\hbar$. Additionally, interpretations or enhancements of IOB involve cooperative processes \cite{Hehlen_94,Hehlen_96,Malyshev_91} and laser heating \cite{Gudel_2000}. In particular, the work \cite{Malyshev_91} pioneered the long series of papers studying local field effects in thin films. Various numerical models using nonlinear energy transfer processes in rare-earth-doped crystals have also been reported to show IOB behavior \cite{Noginov_2003,LiLi_08,LiLi_09}.

Most IOB theories were built to describe either atomic excitation or radiation intensity hysteresis. However, not much attention has been paid to studying spectral properties of the IOB response. In the steady state limit the ``classic'' resonance fluorescence of a still two-level atom in vacuum is known to be a function of atomic excitation or inversion $W$. The line shape of the inelastic component of fluorescent light is determined by three frequencies: the spontaneous decay rate $\Gamma$, the detuning between the laser and the transition frequencies $\Delta$, and the Rabi frequency $\Omega$ \cite{Apanasevich:book,Mollow:book,Mandel:book}. It is apparent that as soon as the atom is placed in a host medium the frequencies take effective values $\bar{\Gamma}$, $\bar{\Delta}$, and $\bar{\Omega}$. Such metamorphose could be detected by an observer. We hypothesize that for an intrinsically bistable system with $W$ hysteresis the spectral patterns would be evidence of which frequency is ``responsible''. This is potentially a good technique to identify the adequate IOB model.

Our work studies IOB as switching between fluorescence spectra and changing of spectral patterns with alternative mechanisms of bistability occurrence. As the first step we will study the original model of IOB based on the Lorentz local-field correction. We suggest a new method for calculations of emission spectra as a convenient and consistent addition to the conventional techniques \cite{Mandel:book}. The equations to describe the non-linear interaction of a two-level atom in the ensemble with the laser field and the properties of the scattered light are derived from the Bogolubov-Born-Green-Kirkwood-Yvon (BBGKY) hierarchy for reduced single particle density matrices of atoms and quantized field modes and their correlation operators. This method is a significant improvement to the atom-photon density operator formalism described in books \cite{Apanasevich:book,Stenholm:book} and numerous papers.

\section{BBGKY and Master Equation}
\label{sec:2}
We consider an ensemble of atomic particles and quantized field modes which we denote by subscript indices $A$ and $F$ respectively. The atoms are assumed to be ``motionless'' so that there is no interaction between them rather than via the field modes. The general structure of the Hamiltonian for such a system is given by
\begin{equation}
  H=\sum_A H_A+\sum_F H_F+\sum_A\sum_F V_{AF}
  \label{Hamiltonian}
\end{equation}
where $H_A$ and $H_F$ are the energies of free particles and $V_{AF}$ describes their binary interactions. In order to introduce our approach in a general manner we do not specify the explicit terms until we finalize the model. The evolution and properties of the system are found from the many-particle total density operator $\rho(t)$ which obeys the von Neumann equation:
\[
i\hbar\dt\rho-[H,\rho]=0.
\]
However, in our study we need to find the solutions for the reduced single-particle operators $\rho_A(t)$ and $\rho_F(t)$. The initial conditions for them can be defined using the fact that if at $t\leq0$ no interaction is observed $\rho(0)$ can be factorized to a product of all single-atom $\rho_A(0)$ and single-photon $\rho_F(0)$ density operators. Further development of the single particle operators as constituents of the ``unfactorable'' system and the steady state mode are to be revealed. This goal can be approached if we substitute the von Neumann equation with the BBGKY hierarchy for reduced density and correlation operators \cite{Akhiezer:Stat,Bonitz:book,Kvasnikov:book}. Some issues of building such an hierarchy for atom-photon systems and proper decoupling of equations were discussed in \cite{GladRoe_1,GladRoe_2}. In this paper we will skip the cumbersome derivations and only point out the main steps to the analyzable system of equations. First of all, let us note that in order to treat the problem properly one needs to consider the actual number of particles and avoid using the thermodynamic limit where the problem is treated in terms of constant densities of material particles. We formally assume all atoms to be individuals and turn to their bulk concentration only when performing spatial integration. Correspondingly, the reduced matrices are obtained from $\rho$ by taking a partial trace, i.e.
\begin{equation}
\rho_{\{S\}}=Tr_{\{S'\neq S\}}\rho,~~Tr_{\{S\}}\rho_{\{S\}}=1.
\label{normcond}
\end{equation}
Here, $\{S\}$ and $\{S'\}$ denote non-overlapping collections of species such that ${\{S'+ S\}}$ is the system described by $\rho$. The reduced multi-particle matrices are substituted with their cluster expansions \cite{Akhiezer:Stat,Bonitz:book}:
\begin{eqnarray}
  \rho_{AF}&=&\rho_A\rho_F+g_{AF},\\
  \nonumber
  \rho_{AFF'}&=&\rho_{A}\rho_{F}\rho_{F'}+g_{AF}\rho_{F'}+g_{AF'}\rho_{F}+g_{FF'}\rho_{A}+g_{AFF'},\\
  \nonumber
  etc.
  \label{Ursel_expansion}
\end{eqnarray}
From this definition it follows that in order to meet \eref{normcond} the trace or even a partial trace of the correlation operator $g_{\{S\}}$ should be strictly zero:
\begin{equation}
 Tr_{\{S\}}g_{\{S\}}=0.
 \label{corrcond}
\end{equation}
With all these starting points the von Neumann equation could be equivalently replaced with an infinite series of coupled kinetic equations. The first two types of equations are precisely
\begin{eqnarray}
  &&i\hbar\dt\da-[\bar{H}_{A},\da]=\sum_{F}Tr_F[V_{AF},g_{AF}]\label{bloch},\\
  &&i\hbar\dt\df-[\bar{H}_{F},\df]=\sum_A Tr_A[V_{AF},g_{AF}].
  \label{wave}
\end{eqnarray}
Here, the l.h.s. terms constitute the von Neumann equations for single-particle operators coupled on the r.h.s. to each other and the remaining system through the atom-field correlations found from  \begin{eqnarray}
  i\hbar\dt g_{AF}-[\bar{H}_{A}+\bar{H}_{F},g_{AF}]
     &=&[\bar{V}_{AF},\rho_{A}\rho_{F}]\\
        \nonumber
     &+&\sum_{F'\neq F}Tr_{F'}[V_{AF'},g_{AFF'}]\label{g_AF}.
 \end{eqnarray}
This equation is written in the limit of the generalized second Born approximation \cite{Bonitz:book,GladRoe_2} for which the contributions from other two-particle correlations are neglected and only partial account for the three-particle correlations is taken:
 \begin{eqnarray}
  \label{g_AFF}
  i\hbar\dt g_{AFF'}&-&[\bar{H}_{A}+\bar{H}_{F}+\bar{H}_{F'},g_{AFF'}]\\
         \nonumber
     &=&[\bar{V}_{AF'},\rho_{F'}g_{AF}].
 \end{eqnarray}
The primes at the particle-reference indices are used to distinguish different particles of the same sort. The higher order correlations are beyond the scope of the current analysis. Equations \eref{bloch}-\eref{g_AFF} are closed and energy conservative. The single particles here are also coupled to the remaining system via the semi-averaged interaction operators or Hartree (mean field) terms in the effective Hamiltonians:
 \begin{eqnarray}
 \label{effect_Hamilton}
 \nonumber
  &&\bar{H}_{A}=H_A+\sum_F\lan V_{AF}\ran_F,\\
  &&\bar{H}_{F}=H_F+\sum_A\lan V_{AF}\ran_A,\\
  \nonumber
  &&\bar{V}_{AF}=V_{AF}-\lan V_{AF}\ran_F-\lan V_{AF}\ran_A.
 \end{eqnarray}
The averages must be understood as $\lan \hat{X}\ran_S=Tr_S(\hat{X}\rho)$. Note that the interaction operator \eref{effect_Hamilton} corrected with the Hartree terms is the one that enables the correlation operators be strictly zero when traced over any correlating particle as stated in \eref{corrcond}.

Now we can specify the light emitters as two-level atoms and for free particles write
 \begin{equation}
  H_A=\hbar\omega_A\hat{\sigma}^z_A
  \label{free_atom},~~
  H_F=\hbar\omega_k\crn\ann,
  \label{free_field}
 \end{equation}
The atom-photon interactions will be treated in the electric-dipole approximation
 \begin{equation}
  V_{AF}=-\pl_A\cdot\E_\bks(\bra_A),
  \label{Hamilton_interaction}
 \end{equation}
for which the dipole moment operator of an atom localized at some point $\bra_A$ is
 \begin{equation}
     \pl_A=\bd^{}_A\al_A+\bd^\ast_A\ar_A
  \label{dipole_oper}
 \end{equation}
and a single mode of the electric field at $\bra$ is given by
 \begin{equation}
    \E_\bks(\bra)=i\sqrt{\frac{2\pi\hbar\omega_k}{\mathbb{V}}}\ee\ann e^{i\bk\cdot\bra}+H.c.
  \label{Quant_field}
 \end{equation}
Here, $\hat{\sigma}^z_A,\hat{\sigma}^\pm_A$ are the population inversion, raising and lowering operators for atoms with transition frequency $\omega_A$ and $\bd_A$ is its dipole transition matrix element. Then $\crn$ and $\ann$ are the creation and annihilation operators that refer to a field mode with frequency $\omega_k$, wave vector $\bk$, $k=\omega_k/c$, and polarization $\ee$. $\mathbb{V}$ denotes the quantization volume. The action of the classic external field is reflected by setting the initial condition for a number of selected modes:
\begin{equation}
   \rho_F(0)=|\alpha_\bks\ran\lan\alpha_\bks|,
   \label{init}
\end{equation}
where $|\alpha_\bks\ran$ indicates a coherent state of the field. We will further use the space--time commutation relation
\begin{equation}
    [\E(\bra,t),\E(\bra',t')]=-i\hbar\G(\bra-\bra',t-t')
\label{comm_rule}
\end{equation}
for the electric field
\[\E(\bra,t)=\sum_\bks\E_\bks(\bra,t),
\;\;\;\sum_\bks \rightarrow\frac{\mathbb{V}}{(2\pi)^3}\int d\bk \sum_s\;\;,
\]
where $\G(\bra-\bra',t-t')$ is the Green's tensor following from the dyadic product of the field operators \cite{Akhiezer:QED,Mandel:book}. At the same time the dyads can be written as
\begin{eqnarray}
    \nonumber    \E(\bra,t)\E(\bra',t')-\hat{\textbf{I}}(\bra,\bra';t,t')=-i\hbar\G^+(\bra-\bra',t-t'),\\
    \E(\bra',t')\E(\bra,t)-\hat{\textbf{I}}(\bra,\bra';t,t')=\;\;\;i\hbar\G^-(\bra-\bra',t-t'),
    \label{time-order}
\end{eqnarray}
for which we introduced the field correlation tensor
\[
    \hat{\textbf{I}}(\bra,\bra';t,t')=\colon\E(\bra,t)\E(\bra',t')\colon
\label{commut_rule}
\]
where $::$ denote normal ordering and $\G^\pm$ are the retarded and advanced Green's tensors respectively. The important averages we use below are
\begin{eqnarray}
\nonumber
\textit{\textbf{I}}(\bra,\bra';t,t')=\lan\hat{\textbf{I}}(\bra,\bra';t,t')\ran,\\
\nonumber
\textit{\textbf{I}}_{inc}(\bra,\bra';t,t')=\lan\lan\hat{\textbf{I}}(\bra,\bra';t,t')\ran\ran\\
=\lan\colon\E(\bra,t)\E(\bra',t')\colon\ran-\lan\E(\bra,t)\ran\lan\E(\bra',t')\ran.
\label{corr_function}
\end{eqnarray}
\Eref{bloch} can be shown to give the master equation if one eliminates its field operators using the standard techniques of quantum optics. First, it is beneficial to transform to the interaction picture for which
\begin{eqnarray}
\nonumber
\tilde{\rho}&=&\exp{\Bigl\{\frac{i}{\hbar}\Bigl(\sum_A H_A+\sum_F H_F\Bigr)t\Bigr\}}\rho\\
&&\exp{\Bigl\{-\frac{i}{\hbar}\Bigl(\sum_A H_A+\sum_F H_F\Bigr)t\Bigr\}}.
\label{int_pict_trans}
\end{eqnarray}
In this picture the remaining Hamiltonian operators \eref{dipole_oper} and \eref{Quant_field} describe interactions and acquire the explicit exponential time dependencies $\pm i\omega_A t$ and $\pm i\omega_k t$. Now we can substitute the formal integrals of \eref{wave} and \eref{g_AF} over time $t'<t$ into \eref{bloch}. Then we assume $\tilde{\rho}(t')\approx\tilde{\rho}(t)$. Retaining the terms which contain $|\bd_A|$ and $|\bd_A|^2$ only and making the use of \eref{comm_rule}-\eref{corr_function} with $\bra'\rightarrow\bra_A$ we come to
 \begin{eqnarray}
 \label{bloch_integ}
  \nonumber
  \dt\tilde{\rho}_A&=&
  \frac{i}{\hbar}\Bigr[\pl_A(t)\cdot\EE_{L}(\bra_A,t),\tilde{\rho}_A\Bigl]\\
  \nonumber
  &+&\frac{i}{\hbar}\int_0^t dt' 
    \bigl[\pl_A(t),\cdot\G^+(0,t-t')\cdot\pl_A(t')\tilde{\rho}_A\bigr]
    \\
   \nonumber
  &+&\frac{i}{\hbar}\int_0^t dt' 
     \bigl[\pl_A(t),\cdot\G^-(0,t-t')\cdot\tilde{\rho}_A\pl_A(t')\bigr]
     \\
  &+&\frac{i}{\hbar}\int_0^t dt' 
   \bigl[\pl_A(t),\bigl[\cdot\textit{\textbf{I}}_{inc}(\bra_A,\bra_A;t,t')\cdot
   \pl_A(t'),\tilde{\rho}_A\bigr]\bigr], 
 \end{eqnarray}
where
\begin{eqnarray}
\nonumber
  \EE_L&=&\EE_0\\
  &+&N\int_{V'}d\bra_{A'}\int_0^t dt'
   \G(\bra_A-\bra_{A'},t-t')\cdot\lan\pl_{A'}(t')\ran
,
\label{local_field_def}
\end{eqnarray}
The first commutator on the r.h.s. of \eref{bloch_integ} describes interaction of atom $A$ with  the local field \eref{local_field_def} in which $\EE_0$ is the external field. It originated properly in accordance with the condition defined in \eref{init}. The integrals in \eref{local_field_def} describe contributions from the other atoms in the ensemble to the resulting effective field acting at position $\bra_A$. Here we used the continuous medium approximation and replaced the summation over atoms with a proper integration over the sample volume $V'$. The prime reflects the property of BBGKY which excludes atom $A$ from the mean field \eref{effect_Hamilton} and eliminates its contribution to \eref{local_field_def}. We will apply the conventional approach to such integration where one should mind an ``empty'' sphere around atom $A$.

The next two integrals in \rf{bloch_integ} describe atomic relaxation due to spontaneous emission. The last term makes account for atomic transitions induced by incoherent photons. However the latter processes are to be neglected in this paper. In order to make a comparison with the ``classic'' works on IOB we restrict our model to the case when there is no reabsorption of scattered light.

The equation for the local field does not make an account for the complete Green's tensor which has the delta-function peculiarity at zero point \cite{GladRoe_1,Vries_98,Fleischhauer_Yelin_99} because this region has been removed from the integration volume. However, if we rewrite \eref{local_field_def} with the full tensor and separate integration over volumes $V$ and $\delta V=V-V'$ the local field becomes easier to handle with for homogenous and isotropic media:
 \begin{eqnarray}
\nonumber
  \EE_L&=&\EE_0\\
  &+&\int_0^t dt'\int_{V} d\bra_{A'}\G(\bra_A-\bra_{A'},t-t')\cdot
  {\bf P}(t')+\frac{4\pi}{3}{\bf P}
 \label{field_final}
,
\label{local_field_fin}
\end{eqnarray}
where we introduced macroscopic polarization of the sample ${\bf P}=N\lan\pl_{A'}\ran$. \Eref{field_final} is exactly the well-known Lorentz local field condition \eref{LFC} which describes the difference between the macroscopic field in a medium and the field acting on each atom therein. This is the first result of this paper which demonstrates that the master equation making an account for the local-field correction can be self-consistently obtained from BBGKY.
\section{Bistability and emission spectra}
\label{sec:3}
The master equation obtained in \sref{sec:2} can now be rewritten in a more transparent form. Without its last term \eref{bloch_integ} becomes independent and may be easily transformed to the IOB equation using the Born-Markov and the rotating wave approximations. Changing back to matrix $\rho$ in \eref{int_pict_trans} and operating in the frame rotating with the laser   frequency $\omega_L$ we get
 \begin{eqnarray}
 \nonumber
   \dt\da=&-&i[\Delta\sigma^z,\da]+i[\ar\bar{\Omega}^\ast+\al\bar{\Omega},\da]\\
  &-&\frac{1}{2}\Gamma\{[\ar,\al\da]-[\al,\da\ar]\},
   \label{master_bist}
 \end{eqnarray}
where $\Delta$ is the detuning, $\bar{\Omega}$ is the effective Rabi frequency and $\Gamma$ is the spontaneous decay rate. Since all operators and frequencies are associated with a single arbitrary atom we may now drop the unnecessary subscripts to simplify the notation. Given the model described above these parameters are exactly the following: $\Delta=\omega_A-\omega_L$, $\bar{\Omega}=\mathbf{d}\cdot\EE_{L}/\hbar$, and $\Gamma=2/\hbar~\bd\cdot \mathrm{Im}\G(0,\omega_A)\cdot\bd=(4/3\hbar)|\bd|^2 k^3$. The contribution of the Lamb shift is neglected.

\begin{figure}[h]
 \begin{center}
  \resizebox{0.8\columnwidth}{!}{
  \includegraphics{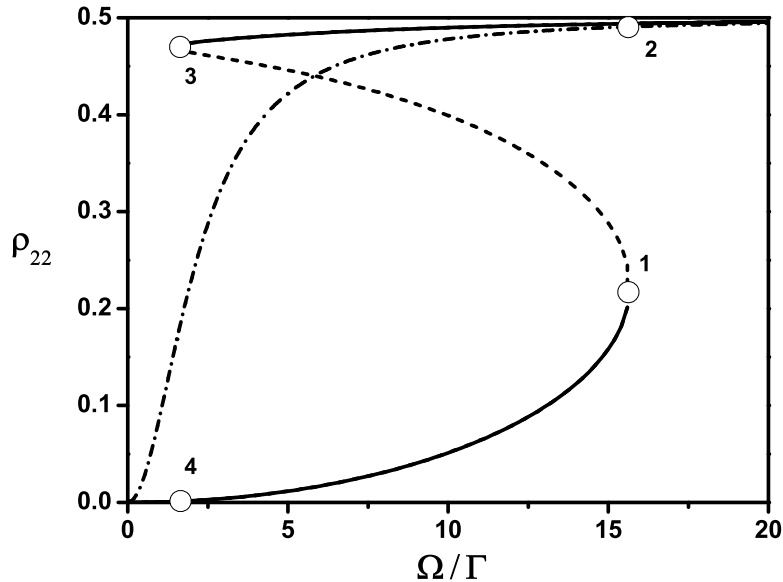}}
 \end{center}
\caption{Excitation hysteresis as function of the laser field strength for $\Delta=3\Gamma$ and $\zeta=50\Gamma$. Solid lines indicate the steady state solutions, dotted line is the unstable branch. The dash-dot line describes the saturation curve of a single independent two-level atom. The laser field strength is expressed in terms of the Rabi frequency $\Omega$ in $\Gamma$ units, $\rho_{22}$ - population density of exited atomic level.}
\label{fig:rho_22}
\end{figure}

This form of the master equation has been very well studied in the steady state limit and is known to describe IOB (see \sref{intro}). When \eref{master_bist} is written as the set of equations for  $\rho_A$ matrix elements and resolved for the population difference function $W=-2\lan\sigma^z\ran$ it produces the third-order algebraic equation:
\begin{equation}
\label{eq:cubic}
c_3 W^3+c_2 W^2+c_1W + c_0 = 0,
\end{equation}
where the polynomial coefficients are:
\begin{eqnarray}
\nonumber &&c_3=\zeta^2,\\
\nonumber &&c_2 = -\zeta(\zeta+2\Delta),\\
\nonumber &&c_1 = \left(2|\Omega|^2 + \Delta(\Delta + 2\zeta)+\frac{1}{4}\Gamma^2 \right),\\
\nonumber &&c_0 = -\left( \Delta^2 + \frac{1}{4}\Gamma^2 \right).
\end{eqnarray}
Coefficient $\zeta$ in our case is precisely the NDD parameter that follows from \eref{local_field_fin}. Consequently, we are to write $\bar{\Omega}=\Omega+\zeta\lan\ar\ran$, where $\Omega=\mathbf{d}\cdot\EE_0/\hbar$ and $\zeta=\zeta_L=(4\pi/3\hbar)N|\mathbf{d}|^2=\pi N\Gamma/k^3$. Three real roots of \eref{eq:cubic} found for various laser field strengths ($\sim\Omega$) form the hysteresis loop of atomic excitation $\rho_{22}=(1-W)/2$ as shown in \fref{fig:rho_22}. Similarly, it is also the function which describes IOB in terms of the total or frequency-integrated power of fluorescent light as $A^{int}\sim\Gamma\rho_{22}$. The coexisting steady state solutions are notated as $\rho_{22}^{low}$ and $\rho_{22}^{up}$ and refer to the lower and the upper curves respectively.

\Eref{eq:cubic} can as well be derived from \eref{master_bist} in which $\Delta\rightarrow\bar{\Delta}=\Delta -\zeta W$ and $\bar{\Omega}=\Omega$. As discussed earlier it could be, for example, either the case suggested in \cite{Messina_2003,Messina_2004} or in \cite{ZhuLi_2000}. In this paper we will not reproduce those models using BBGKY formalism as it would require chains of cumbersome transformations similar to those outlined in \sref{sec:2}. With no new features revealed it could be reasonable to introduce the effective detuning phenomenologically. In this approach we assume $\zeta=\zeta_m=m\Gamma$, where $m$ is an arbitrary factor. In any case with either effective detuning or Rabi frequency we use \eref{eq:cubic} with $\zeta=\zeta_L=\zeta_m$ to find the source of fluorescent light.

Calculation of fluorescent spectra requires the use of \eref{wave} and \eref{g_AF}. In \cite{GladRoe_1,GladRoe_2} the spectrum was defined as irreversible loss of radiation in a nonconservative system. It was shown that such treatment was necessary to calculate the time-dependent or transient mode spectra correctly. The spectra relevant to the steady-state hysteresis do not demand this kind of special care and can be evaluated within the set of equations \rf{bloch}-\rf{g_AFF} that conserve energy. However in this paper we will follow the approach of \cite{GladRoe_1,GladRoe_2} to demonstrate succession of the research techniques applied for related problems.

Let us note first that elimination of three-particle correlation $g_{AFF'}$ in \eref{g_AF} produces operators similar to those found in r.h.s. of \eref{bloch_integ}. Dropping the terms dependent on the number of photons we get the atomic contribution to the correlation function associated with the Lorentz local field and spontaneous emission. Besides, we will supply \eref{wave} and \eref{g_AF} with the field damping operator:
 \begin{equation}
   \hat{\mathcal{L}}_\bks X =-i\eta_k/2\bigl\{[\crn,\ann X]
   -[\ann,X\crn]\bigr\},
 \label{thermostat}
 \end{equation}
where $\eta_k$ is the loss rate at zero temperature. This damping could be introduced accurately within BBGKY if the system is completed with the field reservoir. However, since it is a well known result we simply write it in. Later on in our calculations the damping rate is considered to be $\eta_k\rightarrow 0$.

Meanwhile we have a complete set of equations to study spectral properties of resonance fluorescence. The power density of a field mode is a function of
 \begin{equation}
   A'_k=\hbar\omega_k\dt\lan\crn\ann\ran=\hbar\omega_k Tr_F\Bigl\{\crn\ann\dt\df\Bigr\}.
 \end{equation}
The time derivative of $\df$ is defined by \eref{wave} with attachment of \eref{thermostat}. Since we are to count photons leaving the system we need to consider the nonconservative term only. The latter is nothing but \eref{thermostat}. It is also necessary to write the opposite sign to become an ``observer''. In such a way we come to the function that describes the power of the emission signal absorbed by an idealized detector:
 \begin{equation}
    A_k=
   -\hbar\omega_k Tr_F\{-i\crn\ann\hat{\mathcal{L}}_\bks\df\}
   =\hbar\omega_k\eta_k\lan\crn\ann\ran.
   \label{fluor_general}
 \end{equation}
It must be understood that the photon counter is characterized by a continuity of $\eta_k$. Now our task is to find $\eta_k\lan\crn\ann\ran$ in the steady state. The proper equation for the number of photons follows straightforwardly from \eref{wave}:
 \begin{eqnarray}
 \nonumber
  \dt\lan\crn\ann\ran&+&\eta_k\lan\crn\ann\ran\\
       \nonumber
    &=&\sum_A\xi_k(\bra_A)\lan\ann\ran\lan\ar\ran\\
    &+&\sum_A Tr_A(\ar g_k^+)+H.c.
    \label{photon_num}
 \end{eqnarray}
where $g^+_k =\xi_k(\bra_A)Tr_F(\ann g_{AF})$. The atom-field coupling is represented here by the coupling constant $\xi_k(\bra)=\sqrt{2\pi\omega_k/\hbar\mathbb{V}}(\bd\cdot\ee)e^{i\bk\cdot\bra}$. In the steady state with $d\lan\crn\ann\ran/dt=0$ evaluation of $A_k$ becomes obvious. Functions $\lan\ann\ran$ and $\lan\crn\ran$ are eliminated naturally using \eref{wave} and \eref{thermostat} transformed to the form
\begin{eqnarray}
  \dt \lan\ann\ran =-i\Bigr(\nu_k-i\frac{\eta_k}{2}\Bigl)\lan\ann\ran + \sum_A\xi_k^\ast(\bra_A)\lan\al\ran.
\end{eqnarray}
Function $g^+_k$ and its conjugate are found from the equations obtained from \eref{g_AF} and \eref{thermostat}:
\begin{eqnarray}
  \nonumber
  &&\dt g^+_k + i\nu_k g^+_k + i[\Delta\sigma_z,g^+_k]-i[\ar\bar{\Omega}^\ast+\al\bar{\Omega},g^+_k]\\
  && -\frac{\Gamma}{2}\{[\ar,\al g^+_k]-[\al,g^+_k\ar]\}=|\xi_k|^2(\al\da -\lan\al\ran\da),
  \label{g_plus}
\end{eqnarray}
where we introduced the scale frequency $\nu_k=\omega_k-\omega_L$ and applied $\eta_k\rightarrow 0$. In derivation of these equations we always assumed that all the approximations used to obtain the master equation \eref{master_bist} were also valid. In the steady state for $N$ light emitters with no account for the spatial part we get
 \begin{equation}
 \eta_k\lan\crn\ann\ran=2N^2\xi^2_k|\rho_{12}|^2\frac{\eta_k/2}{\nu_k^2+\eta_k^2/4}
    +N\Bigr\{(g_k^+)_{12}+H.c.\Bigl\}.
    \label{solved}
 \end{equation}
Finally using
 \begin{eqnarray*}
   \lim_{\kappa\rightarrow
   0}\frac{1}{\pi}\frac{\kappa}{x^2+\kappa^2}=\delta(x)
 \end{eqnarray*}
to substitute \eref{solved} into \eref{fluor_general} the expression for the fluorescence spectrum becomes
 \begin{equation}
   A_k=N^2 F_k|\rho_{12}|^2\delta(\nu_k) + 2N\hbar\omega_k\textrm{Re}\Bigr\{(g_k^+)_{12}\Bigl\}
    \label{spectrum_final}
 \end{equation}
where the off-diagonal matrix elements $\rho_{12}=\lan\ar\ran$ and $(g_k^+)_{12}=Tr_A(\ar g_k^+)$ are yet to be found from \eref{master_bist} and \eref{g_plus} respectively with the time derivatives put to zero. $F_k$ is a power density dimension factor. It is important to note that \eref{g_plus} is linear and can be solved explicitly as shown further. When $\bar{\Omega}=\Omega$ the final expression becomes the famous formula for the fluorescence spectrum with coherent and incoherent parts. However, unlike the method described in \cite{Apanasevich:book,Stenholm:book} we performed no artificial steps to separate the two parts. Therefore, the technique we present is convenient for studying spectral properties of systems exhibiting nonlinear response to external light for which analytical solution of \eref{master_bist} is hard or impossible to find.
\begin{figure}[h]
 \begin{center}
  \resizebox{0.8\columnwidth}{!}{
  \includegraphics{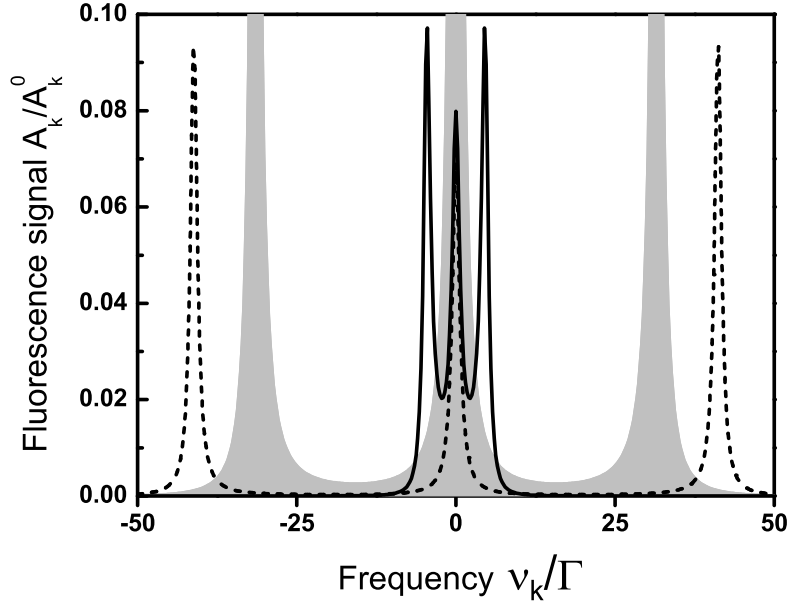}}
 \end{center}
\caption{Emission spectrum at point 1 (lower branch) for $\Delta=3\Gamma$, $\Omega=\Omega^\uparrow\approx15.6\Gamma$. The solid line represents emission spectrum for $\zeta_L=50\Gamma$, $\zeta_m=0$, Dotted line - for $\zeta_m=50\Gamma$, $\zeta_L=0$. The gray shape describes the emission spectrum for an independent two-level atom. $A_k^0$ is maximum value of the emission power function for a single independent atom in the saturation limit.}
\label{fig:fl_1}
\end{figure}

\section{Comparison of spectral patterns}
\label{sec:4}
As we have already pointed out the equations obtained in the previous section are suitable to describe two mechanisms of bistability, i.e., IOB produced by either $\bar{\Omega}$ or $\bar{\Delta}$. The latter model will give \eref{g_plus} effective detuning $\bar{\Delta}=\Delta - \zeta_m W$ and $\bar{\Omega}=\Omega$. In this section we will show the differences in spectral patterns of inelastic fluorescent emission expected for $\bar{\Omega}$ and $\bar{\Delta}$ based IOB models given $\zeta_L=\zeta_m$ and excitation curves are identical as plotted in \fref{fig:rho_22}. We can perform a formal generalization of \eref{g_plus} by writing $\Omega$ and $\Delta$ and deciding which is to represent an effective value later in the final expression. From \eref{spectrum_final} it follows that we need the function of $\textrm{Re}\Bigr\{(g_k^+)_{12}\Bigl\}$ which can be found from \eref{g_plus}. It can be expanded to the system of four equations for four matrix elements $(g_k^+)_{11}, (g_k^+)_{12}, (g_k^+)_{21}, (g_k^+)_{22}$. However, we can reduce the number of equations to three using the general property of a correlation operator \eref{corrcond} which in this case gives $(g_k^+)_{11}+(g_k^+)_{22}=0$. Consequently, the set of equations can be represented in the matrix form as follows:
 \begin{equation}
   M\times g = q,
   \label{equation}
 \end{equation}
where M is the matrix:
 \begin{equation}
   M=\left(
   \begin{array}{ccc}
    i\nu_k + \Gamma & i\Omega^\ast                        & -i\Omega\\&&\\
    2i\Omega        & i(\nu_k-\Delta) + \frac{1}{2}\Gamma & 0\\&&\\
    -2i\Omega^\ast  & 0                                   & i(\nu_k+\Delta) + \frac{1}{2}\Gamma
   \end{array}
   \right),
 \label{Matrix_M}
 \end{equation}
and $g$ and $q$ are vectors. The first represents the correlation function in terms of its matrix elements
 \begin{equation}
   g=
   \left(
   \begin{array}{ccc}
    (g_k^+)_{11},~
    (g_k^+)_{12},~
    (g_k^+)_{21}~
   \end{array}
   \right)^T.
 \label{Matrix_3}
 \end{equation}
Vector $q = |\xi_k|^2 s - |\xi_k|^2\rho_{21}\rho_A$ has two constituents where $s$
represents the source of spontaneously emitted photons
 \begin{equation}
   s=
   \left(
   \begin{array}{ccc}
    \rho_{21},~
    \rho_{22},~
    0
   \end{array}
   \right)^T
 \label{Matrix_S_1}
 \end{equation}
and vector $\rho_A$ eliminates the Rayleigh singularity
 \begin{equation}
  \rho_A=
   \left(
   \begin{array}{ccc}
    \rho_{11},~
    \rho_{12},~
    \rho_{21}
   \end{array}
   \right)^T.
 \label{Matrix_rho}
 \end{equation}
Solution of \eref{equation} gives
\begin{equation}
(g_k^+)_{12} = -2\rho^2_{22} |\xi_k|^2\left(\frac{\nu_k^2 - \Gamma^2 - 2|\Omega|^2 - 2i\Gamma\nu_k}{\det M}\right).
\end{equation}
In order to obtain the desired expression for $\textrm{Re}\Bigr\{(g_k^+)_{12}\Bigl\}$ let us multiply the numerator and denominator by the complex conjugate term $\det M^\ast$ and take the real part of the numerator:
\begin{equation}
 \textrm{Re}\Bigr\{(g_k^+)_{12}\Bigl\}=
   2\rho^2_{22}|\xi_k|^2\Gamma a \left(\frac{\nu_k^2 + a_0}{\nu_k^6 + b_4\nu_k^4 + b_2\nu_k^2 + b_0}\right).
\label{main result}
\end{equation}
The corresponding polynomial coefficients are:
\begin{eqnarray}
\nonumber &&  a = 2|\Omega|^2 + \Delta^2 + \frac{1}{4}\Gamma^2\\
\nonumber &&a_0 = 2|\Omega|^2 + \Gamma^2\\
\nonumber &&b_4 = - 8|\Omega|^2 - 2\Delta^2 + \frac{3}{2}\Gamma^2\\
\nonumber &&b_2 = 16|\Omega|^2 + 2|\Omega|^2(4\Delta^2 + \Gamma^2) + \Delta^4
                         - \frac{3}{2}\Gamma^2\Delta^2 + \frac{9}{16}\Gamma^4\\
\nonumber &&b_0 = \Gamma^2\left( 2|\Omega|^2 + \Delta^2 +\frac{1}{4}\Gamma^2\right)^2
\end{eqnarray}
\Eref{main result} with proper substitutes for either $\Omega$ or $\Delta$ is the main results of our paper. It determines the shapes of the inelastic emission spectra in a wide range of parameters. Below we present several plots of the spectrum profiles corresponding to four important points on the hysteresis curve in \fref{fig:rho_22}. Points $1$ and $2$ give the atomic excitations around the threshold value of $\Omega=\Omega^\uparrow$ that refers to the switch from the lower to the upper branch of the hysteresis loop. For a single two-level atom in the steady state one cannot expect the excitation to be as low as indicated by point $1$ in the limit of a strong field. The threshold point for the lower curve is sufficiently far in the saturation region. Consequently, the spectral function \eref{main result} for $\rho_{22}=\rho^{low}_{22}(\Omega^\uparrow)$ is notably different from conventional triplets \cite{Mollow:book,Mandel:book}. \Fref{fig:fl_1} demonstrates the plots for $\bar{\Omega}$ and $\bar{\Delta}$ variants of IOB. One can see that the line splittings and peak values are sharply ``outlined'' by the respective models. The distance between the satellites and the central peak is over $10$ times greater in terms of $\Gamma$ units for IOB response conditioned by effective $\bar{\Delta}$ as compared to the Lorentz model.  Moreover at this point the linear response spectrum is sharply distinctive as it is nothing but the ``classic'' Mollow's profile.
\begin{figure}[h]
 \begin{center}
  \resizebox{0.8\columnwidth}{!}{
  \includegraphics{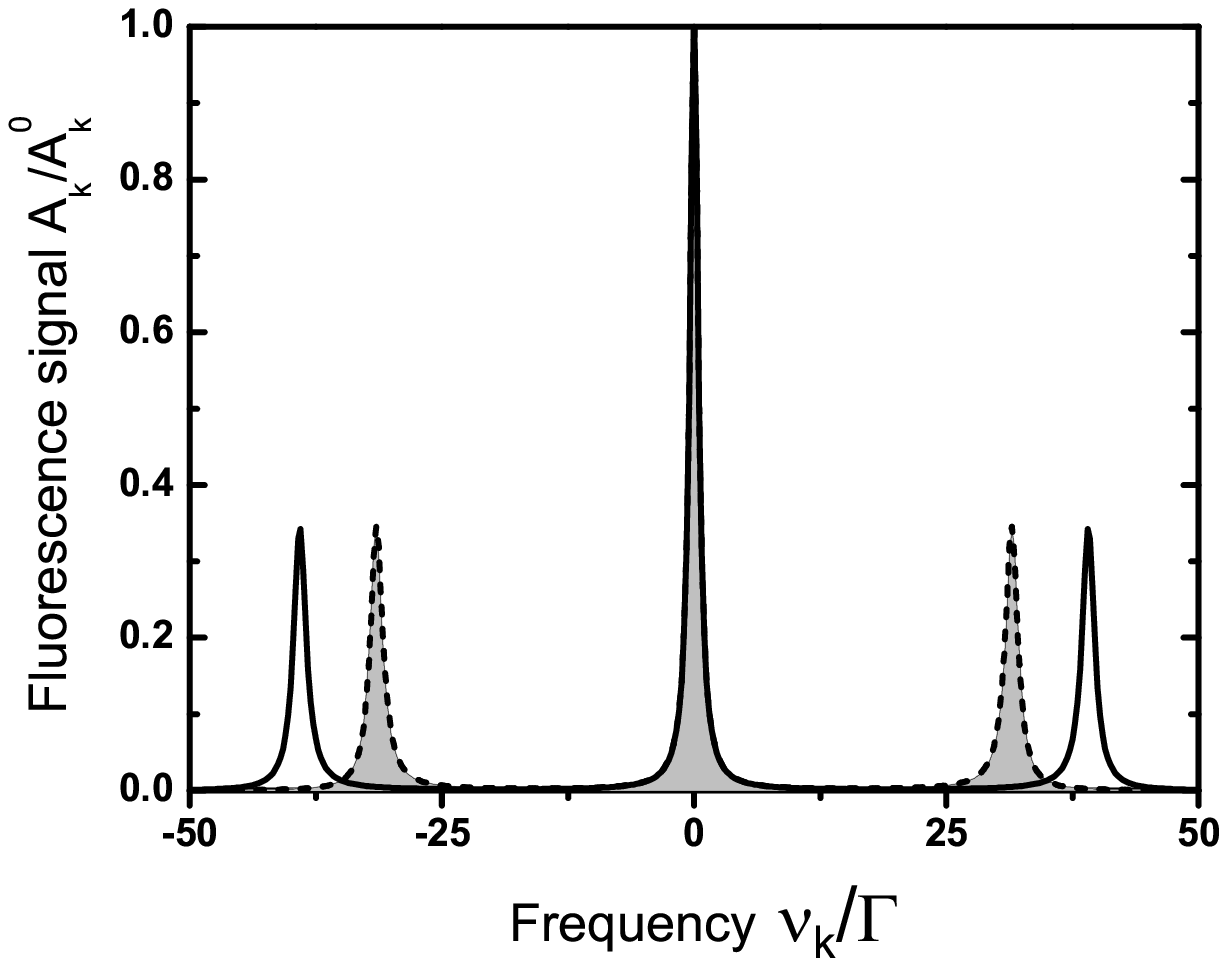}}
 \end{center}
\caption{Emission spectrum at point 2 (upper branch) for $\Delta=3\Gamma$, $\Omega=\Omega^\uparrow\approx15.6\Gamma$. Solid line represents emission spectrum for $\zeta_L=50\Gamma$, $\zeta_m=0$, dotted line - for $\zeta_m=50\Gamma$, $\zeta_L=0$.}
\label{fig:fl_2}
\end{figure}

For the upper branch at point 2 we have $\rho_{22}=\rho^{up}_{22}(\Omega^\uparrow)$ and the emission spectra plots are shown in \fref{fig:fl_2}. Since here $W\approx0$, the contribution from the effective detuning is weak, i.e. $\bar{\Delta}\approx\Delta$, and $\Omega>>\Delta, \Gamma$. Obviously, the respective triplet coincides with the classic shape. In its turn the Lorentz IOB spectrum still reflects some difference between $\Omega$ and $\bar{\Omega}$. However, it will gradually vanish with further increase of $\Omega$. This can be very well seen from the strict relation
\begin{equation}
   |\bar{\Omega}|^2=|\Omega|^2\frac{\bar\Delta^2+\Gamma^2/4}
   {(\bar\Delta-\zeta_L W)^2+\Gamma^2/4},
   \label{Rabi_effect_rel}
\end{equation}
that follows from \eref{master_bist}. As one can see in \eref{main result} the line shape depends on $|\bar{\Omega}|^2$ and the relation to the spectral properties of a single atom is determined by \eref{Rabi_effect_rel}.

\begin{figure}[h]
 \begin{center}
  \resizebox{0.8\columnwidth}{!}{
  \includegraphics{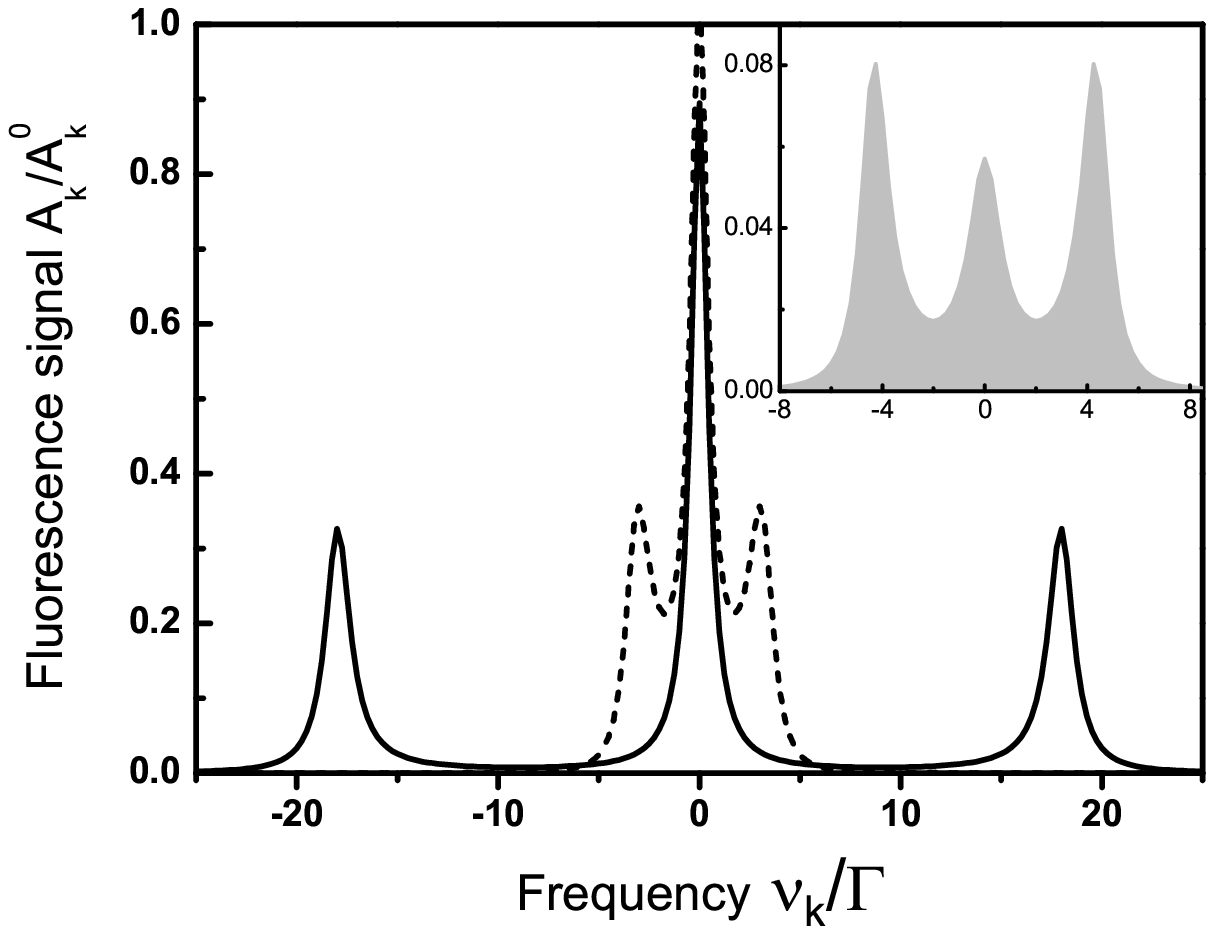} }
 \end{center}
\caption{Emission spectrum at point 3 (upper branch) for $\Delta=3\Gamma$, $\Omega=\Omega^\downarrow\approx1.6\Gamma$. Solid line represents emission spectrum for $\zeta_L=50\Gamma$, $\zeta_m=0$, dotted line - for $\zeta_m=50\Gamma$, $\zeta_L=0$.}
\label{fig:fl_3}
\end{figure}

As we reduce the strength of excitation we move to point $3$ on the upper steady state curve. It corresponds to the weak threshold field $\Omega^\downarrow$, high excitation and small $W$. For these parameters the emission spectra are shown in \fref{fig:fl_3}. Here we find smaller splitting between spectral lines as compared to the single atom picture. Opposite to point 1 the $\bar{\Delta}$ bistability triplet has a narrower range in comparison with $\bar{\Omega}$ IOB.

Further reduction of the pump strength will make the system transit to point $4$ at which $W\approx1$. Since $A_k$ is proportional to $\{\rho^{low}_{22}(\Omega^\downarrow)\}^2$ the signal intensity should be very small and disadvantageous for registering the line shapes.

\begin{figure}[h]
 \begin{center}
  \resizebox{0.75\columnwidth}{!}{
  \includegraphics{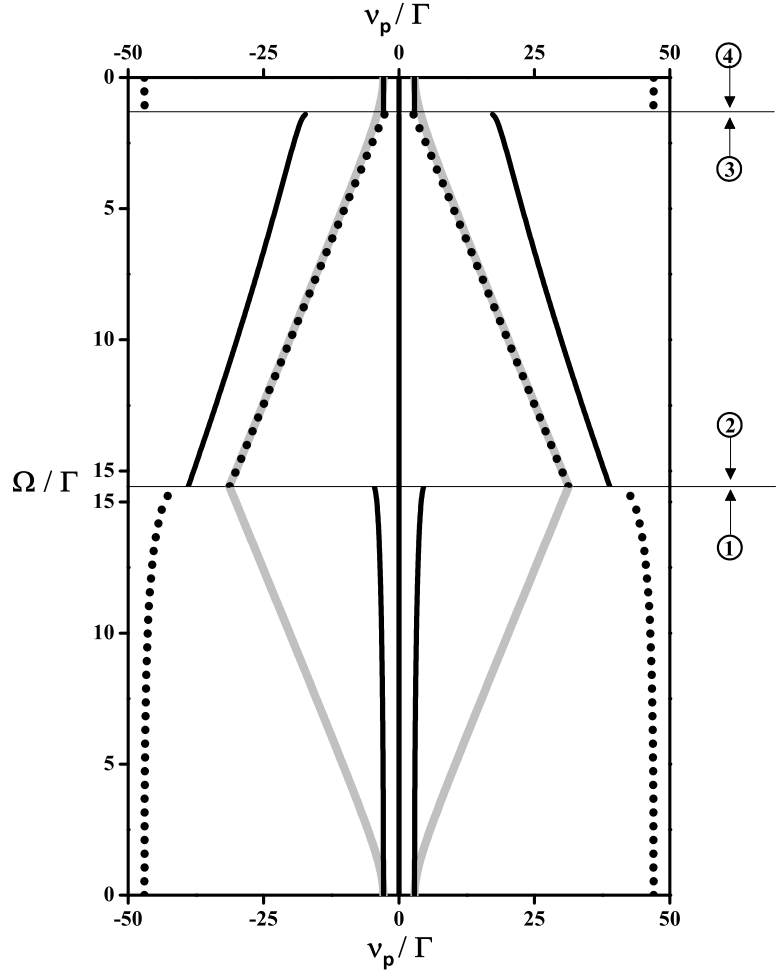} }
 \end{center}
\caption{Positions of emission spectra peaks for the lower and upper branches of the hysteresis loop. Solid line represents peaks position due to local field correction $\zeta_L=50\Gamma$, $\zeta_m=0$, dotted line - due to effective detuning $\zeta_m=50\Gamma$, $\zeta_L=0$. Gray solid lines describe positions of spectrum peaks of an single independent two-level atom with a laser field applied in the vacuum.}
\label{fig:peaks}
\end{figure}

Let us consider the denominator in equation \eref{main result} for the emission spectrum:
\begin{equation}
\frac{1}{\nu^6_k + b_4\nu^4_k + b_2\nu^2_k + b_0}
\end{equation}
It can be written as a product of the peak roots and two positive terms in the following form:
\begin{equation}
\label{eq:peaks}
\frac{1}{\nu^2_k(\nu_k-\nu_p)^2(\nu_k+\nu_p)^2 + 8\Gamma^2|\bar\Omega|^2\nu^2_k + \gamma^6}
\end{equation}
where
\begin{equation}
\label{eq:nu_p}
\nu_p=\sqrt{4|\bar\Omega|^2 + \bar\Delta^2 -\frac{3}{4}\Gamma^2}
\end{equation}
and
\begin{equation}
\gamma^6=\Gamma^2\left(2|\bar\Omega|^2 + \bar\Delta^2 + \frac{\Gamma^2}{4}\right)^2
\end{equation}
From \eref{eq:peaks} the triplet structure of fluorescence becomes clear. The central peak at $\nu_k=0$ is symmetrically surrounded by the satellites with maximums at $\nu_k = \pm\nu_p$. It follows straightforwardly from the fact that if $(\nu_k\pm\nu_p)^2=0$ function \eref{main result} reaches its maximum value. In addition to the conventional analysis of fluorescence these expressions contain the effective or renormalized frequencies. \Eref{eq:nu_p} gives the exact  position of the side peaks for both models of IOB. One may write the proper substitute for $|\bar\Omega|^2$ or $\bar\Delta^2$ to introduce the respective mechanism and obtain a picture similar to the one shown in \fref{fig:peaks}. It is demonstrated here that the frequency shifts of the satellites in a strong fluorescence signal may vividly characterize which mechanism is likely to produce a saturation curve with an IOB hysteresis loop. Moreover, \eref{eq:nu_p} is also suitable for studying joint contribution of effective parameters. However, this issue as well as calculation of $\rho_{22}$ to find a complex IOB picture is out of discussion in this paper. Finally, we may note that the Bowden-Lorentz type of IOB is best detected at point $3$ or, alternatively, on the lower branch as narrowing of the spectral range. At the same time the real renormalization of the atomic transition frequency could be very well seen when uprising to the threshold point $1$.

\section{Summary}
\label{sec:5}
We have performed a purely analytical study of IOB as switching between the patterns of fluorescence spectra. The local field induced hysteresis was shown to originate naturally for the model of a dense collection of like emitters when described thoroughly within the BBGKY hierarchy. The steady state spectra were found self-consistently within the developed formalism using the atom-field correlation matrix. We demonstrated that in terms of spectral analysis the contribution from the Lorentz local field cannot be described via introduction of an effective excitation depended detuning. In other words no evidence of actual renormalization of the atomic transition frequency can follow from observing light scattering in a monocomponent system. Within this concept the incoherent Mollow triplets would reflect abnormally small effective pumping of light emitters along the lower branch of the hysteresis loop. At the same time, a significant increase of the field induced splitting must be expected for the high excitation state. The latter though should gradually disappear in the saturation limit.

In order to make a comparison with the models suggesting renormalization of the transition frequency due to interactions of the atom with its complex environment we introduced this property phenomenologically. Since the IOB equation for atomic excitation is valid for both concepts the system parameters were chosen to give identical hysteresis loops. Analysis of the spectral profiles for this case revealed substantial differences in the character of line splitting at weak and moderate excitation. The results of our comparison are supported by analytical equations \eref{main result}-\eref{Rabi_effect_rel} and \eref{eq:nu_p}.

This analysis was carried out to emphasize the importance of spectroscopic data in studying optical response of complex materials. The problem could be described theoretically with the BBGKY formalism adjusted for atom-photon interactions. It is very advantageous to use this approach because the theory could be created from the first principles avoiding common phenomenology. The BBGKY method applied in this paper gave us a self-consistent set of equations with remarkable features which revealed some optical properties of the atomic system and the scattered field. It was due to the BBGKY concept and the general property of correlation function that we succeeded in reducing the number of equations to obtain the fluorescence spectrum function. The Rayleigh elastic constituent of the scattered light was automatically separated from the broad inelastic components which simplified our calculations greatly as compared to other spectroscopic techniques using the density matrix. The formula obtained for emission spectra can be used to distinguish among possible contributions from effective atomic frequencies in a medium showing IOB in experiments.
\section*{References}
\bibliography{MGladBib}
\end{document}